\def\tipo{2}

\ifnum \tipo = 2
 \documentstyle[prl,twocolumn,aps,psfig]{revtex}
 \def\figsize{8cm}
 \def\figsiz1{8cm}
 \def \frontmatter{\twocolumn[\hsize\textwidth\columnwidth\hsize\csname@twocolumnfalse\endcsname}
 
\else
 \documentstyle[preprint,aps,prl,psfig]{revtex}
 \def\figsize{12cm}
 \def\figsiz1{12cm}
 \def\frontmatter{}
 
\fi

\begin{document}
%\preprint{}
\draft
\frontmatter
\title{ 
A Stochastic Model of Human Gait Dynamics}
\author{Yosef Ashkenazy$^{1}$, Jeffrey M. Hausdorff$^{2}$,
Plamen Ch. Ivanov$^{1,2}$,
Ary L. Goldberger$^{2}$, and H. Eugene Stanley$^{1}$}
\address{ 
$^1$ {\it Center for Polymer Studies and Department of Physics, Boston
University, Boston, Massachusetts 02215, USA}\\
$^2$ {\it Beth Israel Deaconess Medical Center,
Harvard Medical School, Boston, Massachusetts 02215, USA}
}
\date{\today}
\maketitle
\begin{abstract}
{ 
We present a stochastic model of gait rhythm dynamics, based on
transitions between different ``neural centers'', that reproduces
distinctive statistical properties 
of normal human walking. By tuning one model parameter, the
hopping range, the model can describe alterations in
gait dynamics from childhood to adulthood --- including a decrease in the
correlation and volatility exponents with maturation. The model also
generates time series with multifractal spectra whose broadness depends
only on this parameter.
}
\end{abstract}
%02.50.Ey Stochastic processes
%05.40.-a Fluctuation phenomena, random processes, noise, and Brownian motion
%87.90.+y Other topics in biological and medical physics
%87.15.Aa Theory and modeling; computer simulation
\ifnum \tipo = 2
]
\fi

\def\figureI{
\begin{figure}[thb]
\centerline{\psfig{figure=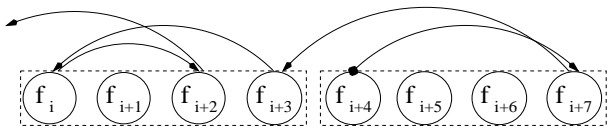,width=\figsize}}
{
\ifnum\tipo=2
\vspace*{0.0truecm}
\fi
\caption{\label{fig1} 
Illustration of the ``neural'' hopping model. 
The values of $f_i$ are not uncorrelated but rather have a finite
size correlations. Shown is a sequence of four transitions, from
mode $f_{i+4}$ to $f_{i+7}$ to $f_{i+3}$ to $f_i$ to $f_{i+2}$ ...
Larger values of the hopping-range parameter $C$ are associated with
larger ``jump sizes'' along the chain.
The neuronal zone of size $\delta_0=4$ is indicated by the dashed boxes.
}}
\end{figure}
}

\def\figureII{
\begin{figure}[thb]
\centerline{\psfig{figure=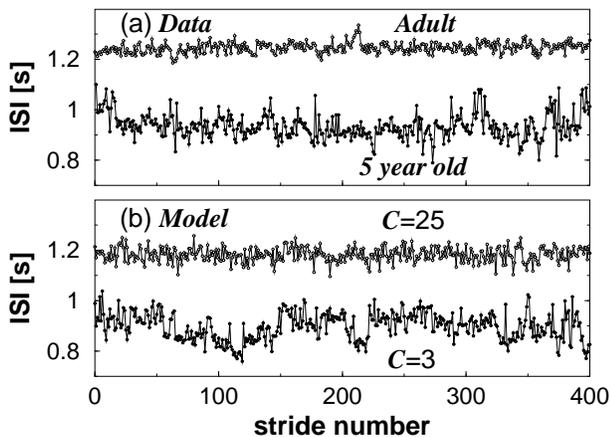,width=\figsize,angle=-90}}
{
\ifnum\tipo=2
\vspace*{0.0truecm}
\fi
\caption{\label{fig2} 
(a) Examples of ISI series of healthy subjects, ages 5 and 25 years.   
(b) Examples of ISI generated by the model. Iterating the model with a
small value of the hopping-range parameter ($C=3$) mimics the ISI of young
children, while a large value ($C=25$) mimics that of adults.
}}
\end{figure}
}

\def\figureIIb{
\begin{figure}[thb]
\centerline{\psfig{figure=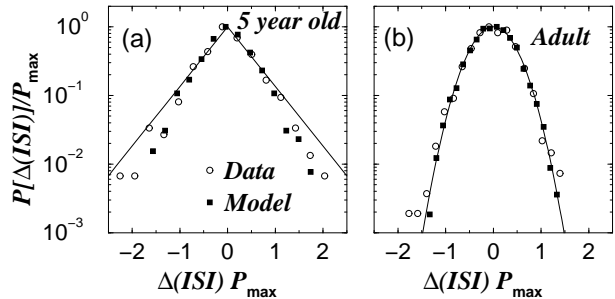,width=\figsize,angle=-90}}
{
\ifnum\tipo=2
\vspace*{0.0truecm}
\fi
\caption{\label{fig2b} 
(a) The normalized distribution of the increments of ISI series ($\Delta
(ISI)_i=(ISI)_{i+1}-(ISI)_i$) for the child's ISI series
(Fig. \protect\ref{fig2}a) and for 
the model with $C=3$ (Fig. \protect\ref{fig2}b); both model and data are
consistent with an exponential distribution, $P(x)=e^{-2|x|}$ (solid line).
(b) Same as (a) for the adult shown in Fig. \protect\ref{fig2}a and for
model ($C=25$) shown in Fig. \protect\ref{fig2}b; in this histogram, both
data and model are consistent with a Gaussian distribution,  
$P(x)=e^{-\pi x^2}$ (solid line).
}}
\end{figure}
}

\def\figureIII{
\begin{figure}[thb]
\centerline{\psfig{figure=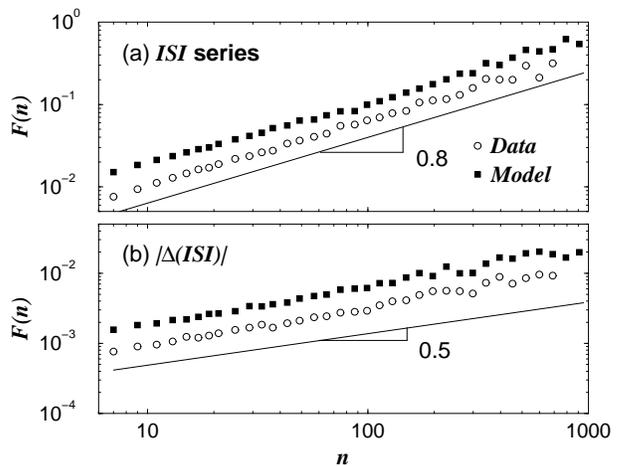,width=\figsize,angle=-90}}
{
\ifnum\tipo=2
\vspace*{0.0truecm}

\fi
\caption{\label{fig3} 
(a) The rms fluctuation function $F(n)$ for the ISI series of
adult (open circles) and of the simulation ($C=25$, black squares) shown 
in Fig. \protect\ref{fig2}. Here $n$ indicates the window size in stride
number. Both the data and the model have similar long-range
correlation properties for the ISI series.
(b) In contrast, $F(n)$ of the magnitudes of the ISI increments $|\Delta
(ISI)_i|$ shows weak correlations (scaling exponent of $\approx
0.55$) where the scaling exponent of an uncorrelated series is $ \approx
0.5$. 
}}
\end{figure}
}

\def\figureIV{
\begin{figure}[thb]
\centerline{\psfig{figure=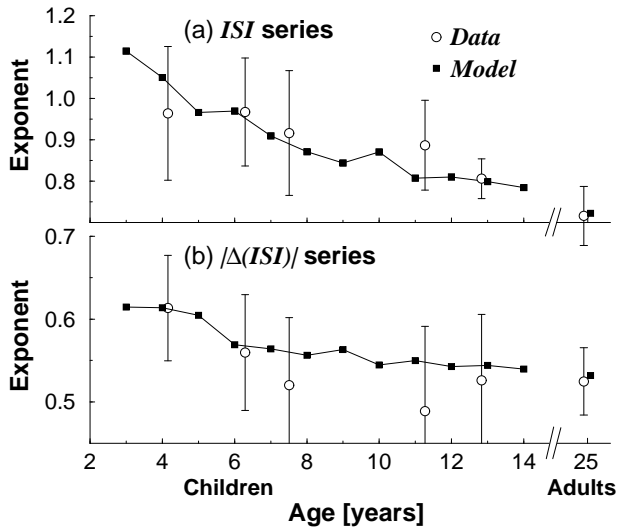,width=\figsize,angle=-90}}
{
\ifnum\tipo=2
\vspace*{0.0truecm}
\fi
\caption{\label{fig4} 
The scaling exponents of the ISI series of the gait maturation database
\protect\cite{MIT} and for the model. The gait maturation
database consists of ISI time series from 50 children between 3 and 13 years
old. Each time series is around 8 minutes ($\sim$ 500 data
points). To study the effects of maturation, we
divide the database into 5 subgroups: (i) 3-4 year olds (11 subjects), (ii)
5-6 year olds (10 subjects), (iii) 7-8 year olds (14 subjects), (iv) 10-11
year olds (10 subjects), and (v) 12-13 year olds (5 subjects). We also show
data \protect\cite{Jeff95} from an adult group (10 subjects 1 hour long
each; ages 20-30 years). 
Values for the scaling exponent axis are given as
mean $\pm$ standard deviation.
For the model simulation, we 
generate 40 realizations for each value of $C$; the average value is
presented. The standard deviation around the average is $\sim$ 0.06 for
the original series and $\sim$ 0.04 for the magnitude series.  The age
axis for the model follows the relation: age (years) $=C+2$.  (a) The
short-range scaling exponents of the original time series both for the
data (open circles) and the model (black squares). The exponents
calculated for window sizes $6 < n < 13$ steps, decrease with age
\protect\cite{Jeff95}. The scaling exponent obtained by the model
decreases monotonically as $C$ increases and is within the error bars of
the data.  (b) The scaling exponent of the ISI magnitude series,
$|\Delta (ISI)_i|$. The magnitude scaling exponent is calculated for
window size $6 < n < 64$ for the children's group and $6 < n < 256$ for
the adult's group; the maximum window size is $\sim {1 \over 10}$ of the
series length for both groups.
The magnitude scaling exponent decreases with age, indicating a loss of
magnitude correlations with maturation. The model exhibits a similar
decrease and the simulation
is within the error bars of the actual data. The subject-to-subject
variability is consistent with the scatter observed in physiologic
indices of neural development \protect\cite{Schroder88}. 
}}
\end{figure}
}

\def\figureV{
\begin{figure}[thb]
\centerline{\psfig{figure=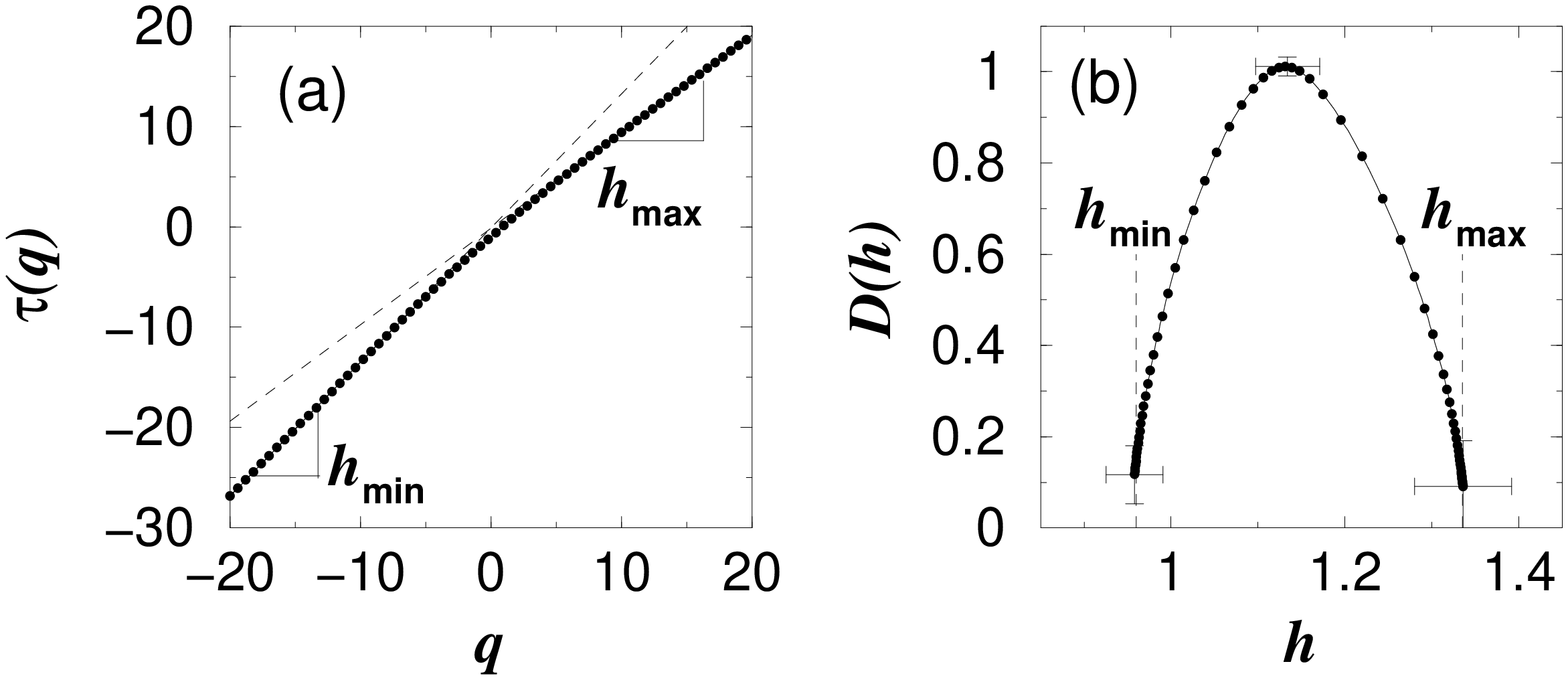,width=\figsiz1,angle=-90}}
\centerline{\psfig{figure=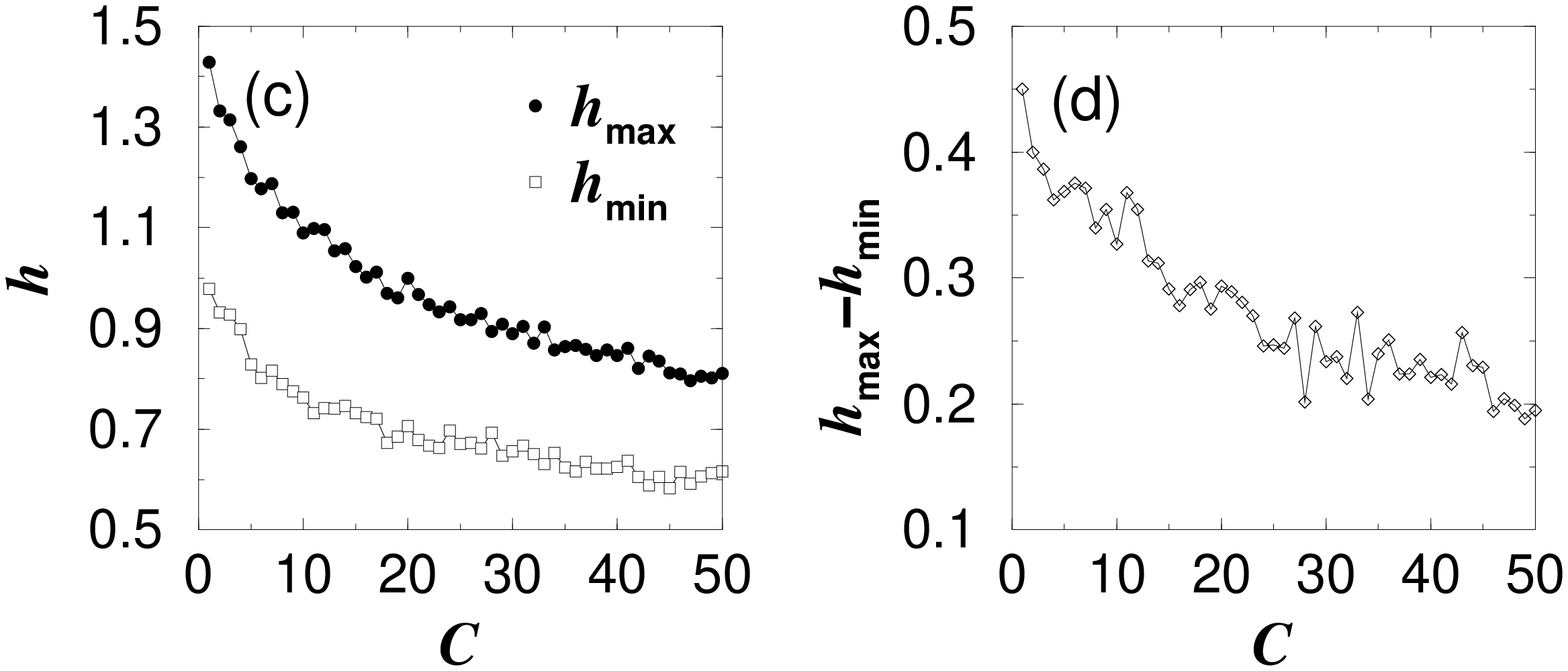,width=\figsiz1,angle=-90}}
{
\ifnum\tipo=2
\vspace*{0.0truecm}
\fi
\caption{\label{fig5} 
The multifractal formalism applied to time series generated by the
model. 
For the multifractal analysis, we applied the wavelet transform modulus
maxima technique \protect\cite{Arneodo} with the Daubechies 10-tap
discrete wavelet \protect\cite{Daubechies}.
We integrate the series before applying the multifractal formalism.
The parameters of the model are fixed (as described in the text)
and only $C$ changes. The series
length is 32768 data points and for each value of $C$, we generate 10
realizations; the average value is presented.
As the first step of the multifractal analysis we extract the partition
function $Z_q(n)$ as a function of the window size $n$ for different
moments $q$. Then we estimate the scaling exponent $\tau(q)$ from the
slope of $\log_2 Z_q(n)$ vs  $\log_2 n$ where $4 \le n \le 2048$.
(a) The exponent $\tau(q)$ vs the moment $q$ for $C=2$. 
(b) The Legendre transform $D(h)=qh-\tau(q)$ of (a) where 
$h(q)=d\tau/dq$ and $D(h)$ is the multifractal spectrum. The maximum and
minimum Holder exponents, $h_{\rm max}$ and $h_{\rm min}$ are 
indicated by dashed lines and estimated as the slope of $\tau(q)$ for
$-20< q \le -15$ and $15 \le q < 20$, respectively (see the
dashed lines in (a)). Error bars (1 standard
deviation) are shown for both axes for $q=-20$, $q=0$, and $q=20$. 
(c) $h_{\rm max}$ and $h_{\rm min}$ for different values of $C$. A
smaller $C$ value yields a broader 
multifractal spectrum. The error bars are as indicated in
(b). (d) The broadness of the multifractal spectrum, $h_{\rm max}-h_{\rm
min}$, as a function of $C$. For larger
$C$, the multifractal spectrum becomes narrower. For $C \gg
1$ $h_{\rm max} \sim h_{\rm min} \sim 0.5$.
}}
\end{figure}
}

Many physical and physiological processes exhibit complex fluctuations
characterized by scaling laws
\cite{Shlesinger,Arneodo95,Ivanov99,Ivanov01,Jeff95,West98,Collins94,Chen97},
some with monofractal 
\cite{Arneodo95} and others with multifractal behavior \cite{Ivanov99}. 
For example, normal gait dynamics of adults are monofractal
\cite{Ivanov01}, while healthy heartbeat fluctuations are multifractal
\cite{Ivanov99}. The origin of these complex fluctuations and the
factors that contribute to differences in their behavior are largely
unknown. Here we develop a physiologically-motivated model
that may be helpful in explaining some of the properties that contribute
to complex dynamics.
We focus on one class of signals ---
the time series of the inter-stride-intervals (ISI) between successive
strides in human gait. 

Walking is a voluntary process, but under normal circumstances, 
stride to stride regulation of gait is controlled by the nervous system
in a largely automatic fashion \cite{Schroder88,BookGait}.
Gait is regulated in part by ``neural centers'' within the cortex
and the spinal cord \cite{BookGait}. As a result of time-varying 
inputs, neural activity at different times are thought to be
dominated by different centers, leading to complex fluctuations
in the ISI ``output.'' Furthermore, as the central nervous system
matures from infancy to adulthood, the interaction between
neural centers becomes richer \cite{Schroder88}.

To understand the underlying regulatory mechanisms of walking, 
deterministic and stochastic models have been proposed. 
For example, classic ``central pattern generator'' models are based on 
oscillatory neural activity, where the interaction 
between neural centers helps regulate gait dynamics
\cite{Collins93}. A stochastic version of a central pattern
generator model reproduces certain fractal properties of
the ISI series \cite{Jeff95}. 
However, existing models do not explain observed changes in scaling
exponents \cite{Jeff95}, and volatility (magnitude)
correlations \cite{remark1,Ashkenazy00} (a new 
finding reported in the present study) that occur during gait maturation
from childhood to adulthood.

We propose a stochastic model consisting of a
random walk (RW) on a chain, the elements of which
represent excitable neural centers \cite{remark2}. 
A step of the RW between element $i$ and element $j$ represents the 
``hopping'' of the excitation from center $i$ to center $j$.
The increase of neural interconnectedness 
with maturation is modeled by increasing the
range of ``jump'' sizes of the RW, since larger jump sizes will allow
exploration of more
neural centers. This property mimics one aspect of the increasing complexity
of the adult nervous system.

Previous studies \cite{Plenz99} have identified neural centers with
pacemaker-like qualities that fire with frequency $f_i$,
so we represent the network of neural centers
by different frequency modes. 
One mode is activated at a given time ($ISI \propto 1/f_i$), and the
$f_i$ are Gaussian distributed. 
The model is based on the following assumptions (Fig. \ref{fig1}) : 

$\bullet$ Assumption (i) is that the $f_i$ have finite-size correlations,
$\langle f_i f_{i+\delta} \rangle / \langle f_i^2 \rangle =
e^{-\delta/\delta_0}$.
We assume finite-range correlations among $f_i$ because
neighboring neurons are likely to be influenced by similar factors
\cite{BookGait}.  
This assumption effectively creates ``neuronal zones'' composed of
neural centers (modes) along the chain with a typical size $\delta_0$.

$\bullet$ Assumption (ii) concerns the rule followed by the RW process.
The active neural center is determined by the location of the RW.
The ``jump'' sizes of the RW follow a Gaussian distribution of width $C$. 

$\bullet$ Assumption (iii) is that a small
fraction of noise is added to the output of each mode  
to mimic biological noise not otherwise modeled. The output $y$ becomes
$y(1+A\eta)$ where $A$ is the noise level and $\eta$ is 
Gaussian white noise with zero mean and unit variance \cite{remark3}.

\ifnum \tipo = 2
\figureI
\figureII
\fi

The model has three parameters $\delta_0$, $C$, and $A$. We find that
the best agreement with the data is 
achieved when $A=0.02$ and $\delta_0=25$. 
In order to simulate changes with maturation, we vary only the third
parameter, $C$, as a function of age, $C=$ (age -- 2) for ages 3 to 25
years. Increasing the hopping range with age is consistent with the fact
that neural transmission is not fully developed until the
late teens \cite{Schroder88,remark5}. 

Examples of ISI time series 
are shown in Fig. \ref{fig2}a. The ISI series of the adult subject has
smaller fluctuations compared with the more variable ISI
of the child. Two examples of the model's output are shown in 
Fig. \ref{fig2}b; using a large value of $C$ ($C=25$) simulates the ISI
series of an adult, while using a small value ($C=3$) mimics the 
ISI series of a child. 
Surprisingly, we find that 
the normalized probability distributions of the child ISI
increment series, $\Delta (ISI)_i=(ISI)_{i+1}-(ISI)_i$, and of the model
with small $C$ are close to exponential (Fig. \ref{fig2b}a). In
contrast, the distribution converges to Gaussian in the adult and for
large $C$ in the model (Fig. \ref{fig2b}b).  

To further test the model, we study time correlations in the ISI
series. We calculate the function $F(n)$, which corresponds to the rms
fluctuations of the integrated ISI series of 
the adult (Fig. \ref{fig2}a) and of the model for $C=25$ (Fig.
\ref{fig2}b) \cite{Jeff95,West98,Peng94}.
Estimating the scaling exponent from the slope of $\log F(n)$ vs $\log n$,
we find, for both data and model, that the ISI time series have
correlations with scaling exponents $\approx 0.8$ (Fig. \ref{fig3}a).   

The scaling exponents of the original signal provide an indication of
the linear 
properties (two-point correlations). Certain nonlinear aspects may be
associated with the presence of long-range correlations in the
magnitudes of ISI increments $|\Delta (ISI)_i|$, an index of
``volatility'' \cite{Ashkenazy00}. We find (Fig. \ref{fig3}b) that the
adult magnitude 
series is uncorrelated (the scaling exponent is close to 0.5).
The model shows similar behavior for large $C$
(Fig. \ref{fig3}b). 

Next we inquire if changes in gait dynamics from childhood to adulthood
might be reflected 
in changes in the scaling exponents of the ISI series.
The short-range scaling exponents of the ISI series
decrease as children mature \cite{Jeff95}. We compare the 
short range scaling exponents for a group of 50 children
\cite{MIT} with those of 10 adults \cite{Jeff95}; this exponent
decreases from $\sim$ 1.0 to $\sim$ 0.7 (Fig. \ref{fig4}a). 
By altering $C$, the model simulates these maturation-related changes in
two-point correlation properties.

\ifnum \tipo = 2
\figureIIb
\fi
\ifnum \tipo = 2
 \figureIII
\fi

The magnitude series exponent of the ISI series also decreases with
maturation (Fig. \ref{fig4}b). 
Our results for the magnitude series suggest that the 
gait pattern of children is more volatile (and thus more nonlinear) than the
\ifnum \tipo = 2
 \figureIV
\fi
usual walking pattern of adults. We note that: (i) unlike adults, 
young children have difficulty keeping their walking speed constant. 
A large ``walking error'' in one direction is likely to be followed by a
large compensatory walking error in the opposite direction --- i.e., a
large (or small) increment of ISI is likely to be followed by a large
(or small) decrement 
over a given range of scales. 
This instability may lead to increased magnitude correlations.
(ii) Adults can voluntarily simulate the less stable gait of children and
thereby increase the magnitude correlations of the ISI. Children,
however, cannot maintain the less volatile dynamics of
adults. 
Thus, in general, the more mature adult gait dynamics are likely to be
richer.  
The model shows (Fig. \ref{fig4}b) a similar
decrease of the magnitude exponents when increasing 
$C$ and is within the error bars of the data.

\ifnum \tipo = 2
 \figureV
\fi
The decrease of the magnitude exponent with maturation is consistent
with the possibility that
the ISI time series (during goal-directed walking) becomes less multifractal
as the individual matures. This prediction cannot be tested on the
available data since the time
series of the children are too short for multifractal
analysis; but this prediction can be tested on the
model. Fig. \ref{fig5} summarizes a systematic multifractal analysis on
ISI series generated by the model. 
We find narrower multifractal spectra with increasing $C$,
suggesting the possibility of a decrease of
multifractality with maturation during normal walking. The model's 
prediction
agrees with a recent finding regarding the monofractality of normal gait
in adults \cite{Ivanov01}. The model also
serves as a generator of random series with different multifractal
spectra --- multifractal for $C \approx 1$ and monofractal for $C \gg
1$. 
 
In summary, we find that a simple stochastic model captures multiple  
aspects of gait dynamics, and their changes with maturation, including:
(i) the shape of the probability distribution of the ISI increments;
(ii) correlation properties of the ISI; and
(iii) correlations properties in the magnitudes of the ISI increments
\cite{remark6}.
Further, by varying only a single ``hopping-range'' parameter, $C$, a
wide array of multifractal dynamics can be generated.
The model can also be altered by ``knocking out'' certain frequency modes
(akin to what may occur during very advanced age or in response to
neurodegenerative disease).  Simulation with drop-out of
frequency modes predicts increased gait variability, with (i) increased
magnitude exponents, and (ii) decrease of long-range correlations. Our
preliminary 
analysis of the ISI series of older adults prone to falls is consistent
with this prediction. 
Generalization of the model to two and three dimensional networks
to describe other types of neurological activities is underway.

Partial support was provided by the NIH/National Center for Research
Resources (P41 RR13622) and the NIA (AG14100). We thank J.J. Collins,
S. Havlin, V. Schulte-Frohlinde, and C.-K Peng for helpful discussions.

\ifnum \tipo = 1
\figureI
\figureII
\figureIIb
\figureIII
\figureIV
\figureV
\fi


\begin{references}
\bibitem{remark2} We model one specific type of human locomotion, namely
``usual'' or ``directed'' gait. This type of voluntary, goal-directed
activity has a large automatic component and usually tends to minimize
variability. Other types of locomotor 
activity (not modeled here) would include more ``free-form'' patterns of
gait, as well as recreational movements during dance or sports.
\bibitem{remark3} It is possible to add the following assumption in
order to capture the observed decrease of ISI variability with age as
children mature \protect\cite{Jeff95}. When in a mode of frequency
$f_i$, the model generates as an output $ISI=B/f_i$. 
We suggest the proportionality constant, $B$, to be the 
ratio between the number of simulated strides and the number of active
modes (i.e., the number of different modes that were visited during a
simulation). 
Then, the standard deviation of the output of the model
decreases when $C$ increases with no effect on the model
dynamics.
\bibitem{remark5} 
The model's dynamics can be understood in the following intuitive way.
For $C\ll \delta_0$ (small hopping-range, young child), the dynamics
will be confined mainly to a single ``neuronal zone''; 
this will lead to correlations in the ISI time series. For $C \approx
\delta_0$  
(large hopping-range, adult) the dynamics will explore frequency modes
from different zones; this will lead to a decrease in the correlations. 
Moreover, for $C\ll \delta_0$ the dynamics changes back and forth from
one mode to a nearby mode leading to increased magnitude
correlations. For $C \approx \delta_0$ the dynamics are governed by many
different modes leading to the loss of magnitude correlations.
\bibitem{remark6} While there is good agreement between the model output
and the data for these orthogonal dynamical properties,
both in adults and as a function of maturation, further study is required
to investigate the physiologic analogs that generate these dynamical
properties.
\end{references}
\end{document}